# Tunable Topological Energy Bands in 2D Dialkali-Metal Monoxides

*Chenqiang Hua, Si Li, Zhu-An Xu, Yi Zheng,\* Shengyuan A. Yang,\* and Yunhao Lu\**

2D materials with nontrivial energy bands are highly desirable for exploring various topological phases of matter, as low dimensionality opens unprecedented opportunities for manipulating the quantum states. Here, it is reported that monolayer (ML) dialkali-metal monoxides, in the well-known 2H-MoS$_2$ type lattice, host multiple symmetry-protected topological phases with emergent fermions, which can be effectively tuned by strain engineering. Based on first-principles calculations, it is found that in the equilibrium state, ML Na$_2$O is a 2D double Weyl semimetal, while ML K$_2$O is a 2D pseudospin-1 metal. These exotic topological states exhibit a range of fascinating effects, including universal optical absorbance, super Klein tunneling, and super collimation effect. By introducing biaxial or uniaxial strain, a series of quantum phase transitions between 2D double Weyl semimetal, 2D Dirac semimetal, 2D pseudospin-1 metal, and semiconductor phases can be realized. The results suggest monolayer dialkali-metal monoxides as a promising platform to explore fascinating physical phenomena associated with novel 2D emergent fermions.

## 1. Introduction

The flourishing of 2D materials and the technical feasibility of fabricating astonishingly complex van der Waals heterostructures[1] allow extremely rich quantum phases of matter to be explored and tuned in atomic thin-films with single-crystal quality. Following the ground-breaking experiments on graphene,[2–4] a variety of archetypal 2D materials have been reported, such as group-(III to VI) elemental 2D layers,[5–8] transition-metal dichalcogenides (TMDCs),[9–12] black phosphorous[13–15] and its binary main group counterparts,[16–18] InSe family materials,[19] and 2D MXenes.[20,21] Very recently, various 2D materials with ferromagnetic and/or ferroelectric ordering have also been explored.[22–28]

In this field, the search for 2D materials with nontrivial topological band structures is an intriguing research topic. It is motivated by recognizing that many unusual properties of graphene are connected to its special topological band structure: the conduction and valence bands linearly cross at symmetry-protected nodal points, such that the low-energy quasiparticles behave like Dirac fermions with zero effective mass. Currently, a target in this direction is to explore new types of band degeneracy points and the associated new emergent fermions. For example, previous works have theoretically proposed 2D massive Dirac fermions,[29] double Weyl fermions, and pseudospin-1 fermions in some 2D materials, such as blue phosphorene oxide[30] and monolayer (ML) Mg$_2$C.[31] However, these proposed materials are not ideal because the emergent fermions are not realized in their equilibrium states. They require finite externally applied strains to be tuned into double Weyl semimetal or the pseudospin-1 metal states, which is

Dr. C. Hua, Prof. Y. Lu
Zhejiang Province Key Laboratory of Quantum Technology
and Device and Department of Physics in Zhejiang University
State Key Lab of Silicon Materials
School of Materials Science and Engineering in Zhejiang University
Hangzhou 310027, P. R. China
E-mail: luyh@zju.edu.cn

Dr. S. Li
School of Physics and Electronics
Hunan Normal University
Changsha, Hunan 410081, China

Dr. S. Li, Prof. S. A. Yang
Research Laboratory for Quantum Materials in Singapore
University of Technology and Design
Singapore 487372, Singapore
E-mail: shengyuan_yang@sutd.edu.sg

Dr. S. Li, Prof. S. A. Yang
Center for Quantum Transport and Thermal Energy Science
School of Physics and Technology in Nanjing Normal University
Nanjing 210023, China

Prof. Z.-A. Xu, Prof. Y. Zheng
Zhejiang Province Key Laboratory of Quantum Technology
and Device and Department of Physics in Zhejiang University
Hangzhou 310027, P. R. China
E-mail: phyzhengyi@zju.edu.cn

Prof. Z.-A. Xu, Prof. Y. Zheng
Collaborative Innovation Centre of Advanced Microstructures
in Nanjing University
Nanjing 210093, P. R. China

challenging to control in experiment. Thus, it still remains as a challenge to search for suitable 2D materials which can host these new fermions.

Meanwhile, among the 2D materials family, the metal-shrouded MXenes have been attracting great interest due to their rich properties.[20,32,33] Most members of the family are carbides and nitrides (e.g., $Ti_2C$, $Ti_2N$, and $Nb_2C$),[33,34] consisting of three atomic layers in the order of M–X–M, where M is the transition metal and X is C or N. Recently, the study has been extended to metal-shrouded oxides.[35] For example, the 2D $Tl_2O$ with the same structure has been proposed. Considering that Tl has properties similar to alkali metal elements, it is natural to ask whether alkali metal oxides (by replacing Tl with alkali metal) can also be stabilized in a 2D crystal form? If so, this may open the door to a new family of 2D materials, i.e., 2D metal-shrouded oxides. In view of the great variety of 2D metal-shrouded MXenes and their rich properties, one can expect that 2D metal-shrouded oxides will also grow into a big family with fascinating properties.

In this work, we try to address the above challenge and question by proposing a new family of 2D material—monolayer dialkali-metal monoxides (DMMOs) with the 2H-$MoS_2$ structure. Based on first-principles calculations, we show that $Na_2O$ and $K_2O$ can be stabilized in this 2D lattice as shown in **Figure 1**a. They exhibit good stability and mechanical property. Most importantly, these materials host a diverse of 2D topological phases with new emergent fermions. In the equilibrium state, ML $Na_2O$ is a 2D double Weyl (massive Dirac) semimetal, while ML $K_2O$ is a nearly ideal 2D pseudospin-1 metal. To our best knowledge, they represent the first examples that host the 2D double Weyl semimetal and the pseudospin-1 metal in the equilibrium state. Furthermore, the topological phase and the novel fermions can be effectively controlled by strain. Under biaxial strain, a series of topological phase transitions between 2D Dirac metal, double Weyl semimetal, pseudospin-1 metal, and semiconductor phases can be realized. Particularly, when uniaxial strain is applied, an intriguing transition from double Weyl point to two single Weyl (Dirac) points is triggered, which would be manifested in quantum transport measurements by a drastic changeover from weak localization to weak antilocalization. This kind of phase transition has not been reported for realistic materials before. In addition, these emergent fermions will lead to a range of exotic effects, including universal optical absorbance, super Klein tunneling, and super collimation effect, which have been predicted but not experimentally demonstrated yet due to the lack of suitable materials. Thus, our discovery not only predicts a family of new 2D materials but also provides an experimentally feasible platform to explore new emergent fermions and their fascinating fundamental physical effects.

## 2. Results and Discussions

The proposed ML dialkali-metal monoxides $A_2O$ (A = Na, K) take the same structure as ML 2H-$MoS_2$, in which three atomic layers are stacked in the A–O–A sequence with the $D_{3h}$ point group symmetry, forming the $P\bar{6}m2$ space group (No. 187). As depicted in Figure 1, the top view of DMMO lattice is honeycomb like, corresponding to a hexagonal first Brillouin zone (BZ). The optimized lattice parameters for $Na_2O$ and $K_2O$ are 3.49 and 3.86 Å, respectively.

We calculated the phonon spectra of DMMOs to investigate the dynamical stability of these compounds. The result for ML $Na_2O$ is shown in Figure 1b. Apparently, there is no imaginary mode in the whole Brillouin zone, indicating that the material is dynamically stable. The phonon spectrum of $K_2O$ is similar to $Na_2O$, as shown in the Supporting Information. Excellent energetic stability has also been validated, as evident by the large cohesive energy of −8.02 and −7.36 eV per formula (f.u.$^{-1}$) for $Na_2O$ and $K_2O$, respectively. These results suggest the experimental feasibility to synthesize the proposed monolayer materials. For the other DMMOs (A = Li, Rb, Cs), we find that $Li_2O$ has some imaginary phonon modes while $Rb_2O$, and $Cs_2O$ is extremely unstable. The calculation results (including their band structures) are summarized in the Supporting Information.

We now focus on the electronic structures of the monolayer DMMOs. In **Figure 2**, we plot the calculated band structures for ML $Na_2O$ and $K_2O$. The band structures of $Li_2O$ are similar to $K_2O$, as summarized in Figure S8 in the Supporting Information. It is noteworthy that, due to the small atomic numbers, spin–orbit coupling (SOC) has negligible effect on the band structure of DMMOs. Consequently, electron spin can be regarded as a dummy degree of freedom and the DMMOs can be treated as effective spinless systems. In the following, unless explicitly stated, we will not take spin degeneracy into account when analyzing the band topology. As shown in Figure 2, both materials are metallic without a bandgap, due to the presence of three energy bands near the Fermi level, originating from the $E'$ and $A_1'$ irreducible representations of the D$_{3h}$ point group (see Figure 2). For Na$_2$O, the quadratic conduction and valence bands touch at a single degenerate point at the high symmetry $\Gamma$ point (Figure 2c). Without external charge doping, such point also defines the Fermi energy due to ionic band filling. The band dispersion around this point is quadratic in all directions in the 2D plane, so this point is named as a 2D double Weyl point similar to that in blue phosphorene oxide and monolayer Mg$_2$C. However, unlike blue phosphorene oxide and Mg$_2$C which both require applied strain to realize such double Weyl point, ML Na$_2$O is intrinsically a 2D double Weyl semimetal. On the other hand, for ML K$_2$O, the three $E'$ and $A_1'$ bands cross each other at the $\Gamma$ point, forming a distinctive triply degenerate point Fermi surface (see the Supporting Information). The difference is mainly due to stronger ionic bonding in ML K$_2$O. Consequently, in the vicinity of triply degenerate point, the upper and lower bands are characterized by graphene-like linear dispersion, intersected by a very flat band at the Fermi point (Figure 2d). Such unique triply degenerate point Fermi surface in ML K$_2$O corresponds to pseudospin-1 fermions, as we will elaborate in the flowing paragraphs.

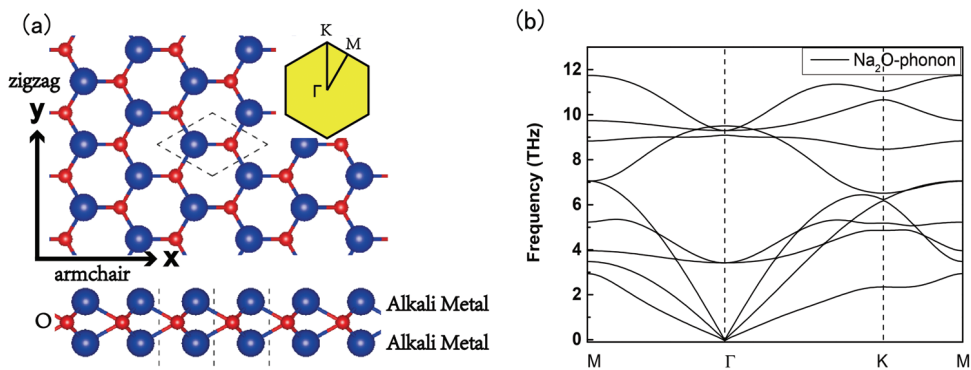

**Figure 1.** a) Top and side view of monolayer dialkali-metal monoxides. Inset is the first Brillouin zone with high-symmetry points. b) Phonon spectrum of ML $Na_2O$, showing no imaginary mode in the whole Brillouin zone.

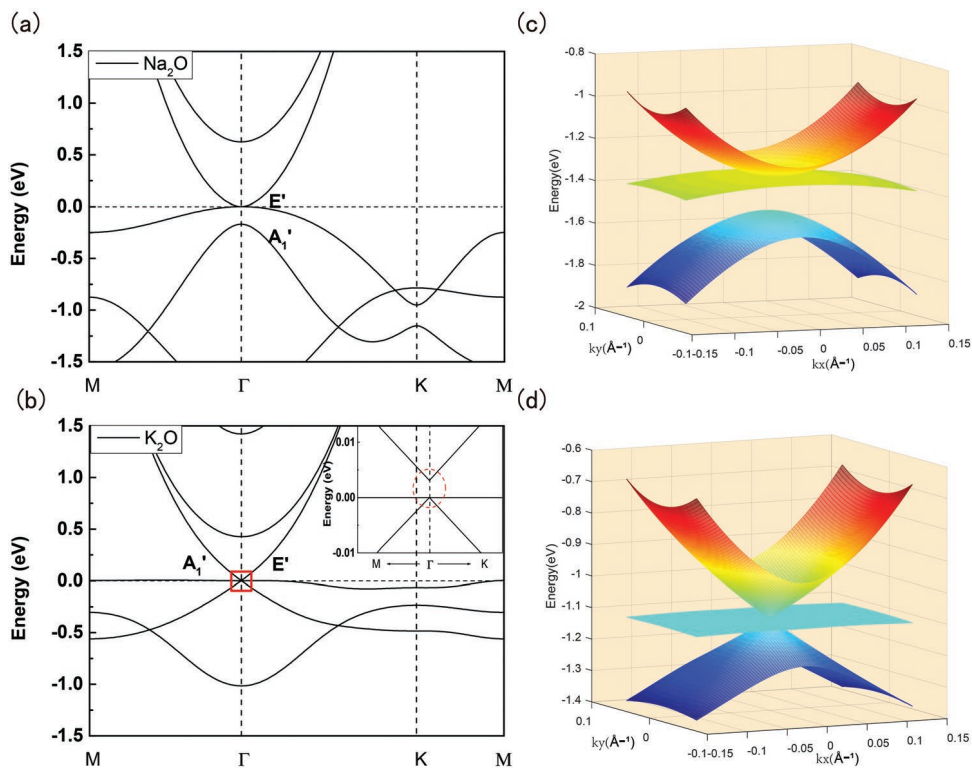

**Figure 2.** a,b) Band structures of ML Na$_2$O and ML K$_2$O, respectively. $E'$ and $A_1'$ are the irreducible representations of the D$_{3h}$ point group, which contribute three bands near Fermi level. The inset of (b) is the zoom-in of the energy bands of K$_2$O at $E_F$ near the $\Gamma$ point. c,d) 2D energy dispersion near the Fermi surface, which are massive Dirac point and triply degenerate pseudospin-1 point for ML Na$_2$O and K$_2$O, respectively.

The triple band degeneracy point and the associated pseudospin-1 fermions can be well modelled by the $\boldsymbol{k}\cdot\boldsymbol{p}$ method, taking the DMMO lattice symmetry constraints into account. Since the band degeneracy points are located at the $\Gamma$ point, we construct the effective model using the $E'$ and $A_1'$ states as bases. Subjected to time reversal symmetry and D$_{3h}$ symmetry operations, which include threefold rotation $C_3$, twofold rotation $C_2$, and horizontal mirror $M_h$, the $\boldsymbol{k}\cdot\boldsymbol{p}$ Hamiltonian expanded up to $k$-quadratic order have the general form of

$$H_0(\boldsymbol{k}) = \begin{bmatrix} M_1 + B_3 k^2 & C(k_y^2 - k_x^2) - iAk_x & 2Ck_x k_y - iAk_y \\ C(k_y^2 - k_x^2) + iAk_x & M_2 + B_1 k_x^2 + B_2 k_y^2 & (B_1 - B_2)k_x k_y \\ 2Ck_x k_y + iAk_y & (B_1 - B_2)k_x k_y & M_2 + B_2 k_x^2 + B_1 k_y^2 \end{bmatrix} \quad (1)$$

where $k = |\boldsymbol{k}|$ is the magnitude of 2D wave vector, $M_{1,2}$, $A$, $B_{1,2,3}$, and $C$ are band parameters which are extracted by fitting the density functional theory (DFT) band structures. Exactly at $\Gamma$ point ($k = 0$), one can see that $M_1$ and $M_2$ represent the energies of the $A_1'$ and the $E'$ states, respectively. By determining the band parameters of this general Hamiltonian, we can quantitatively understand the pronounced changes in the low-energy band structures of Na$_2$O and K$_2$O.

First, for Na$_2$O, $E'$ has higher energy than $A_1'$, meaning $M_1$ is smaller than $M_2$. With a unitary transformation, we can project the $H_0(\boldsymbol{k})$ model onto the two $E'$ bands around the double Weyl (massive Dirac) point. The resulting two-band model is written by

$$H_{DW}(\boldsymbol{k}) = B_+(k_y^2 + k_x^2) + \begin{bmatrix} 0 & B_- k_-^2 \\ B_- k_+^2 & 0 \end{bmatrix} \quad (2)$$

where $B_\pm = 0.5(B_2 \pm B_1)$, $k_\pm = k_x \pm ik_y$, and the energy is measured from the degenerate point. Without the diagonal term, the model $H_{DW}$ resembles that of the AB-stacked bilayer graphene, which features a Berry phase of $2\pi$, doubling that of a single Dirac point as in monolayer graphene.[3] However, for bilayer graphene, there are two such points at BZ corners related by the time reversal symmetry; whereas for ML $Na_2O$, there is only one double Weyl point sitting at the BZ center. The double Weyl point endows ML $Na_2O$ with an interesting optical property, i.e., it should exhibit a universal optical absorbance of $\pi\alpha \approx 2.3\%$ ($\alpha$ is fine-structure constant) at low frequencies [for $\hbar\omega < (M_2 - M_1)$ here].[30,36]

For $K_2O$, the triply degenerate point can be explained by setting $M_2 = M_1 = 0$ (by neglecting the insignificant gap). In this case, the $k$-linear terms dominate the low energy spectrum, which allows us to further simplify the Hamiltonian $H_0$ into the following form

$$H_{PS1}(\boldsymbol{k}) = \begin{bmatrix} 0 & -iAk_x & -iAk_y \\ iAk_x & 0 & 0 \\ iAk_y & 0 & 0 \end{bmatrix} = A\boldsymbol{k} \cdot \boldsymbol{S} \quad (3)$$

where $\boldsymbol{S}$ is pseudospin-1 angular momentum vector.[37] Although $H_{PS1}$ mimics the Hamiltonian of 3D Weyl, each $\boldsymbol{S}$ matrix is $3 \times 3$, representing the triply degeneracy of pseudospin-1 fermions in $K_2O$ near the Fermi surface. In contrast to blue phosphorene oxide and $Mg_2C$, the middle intersecting band is extremely flat, nearly extending over the whole Brillouin zone. Intriguingly, ML $K_2O$ exhibits pseudospin-1 phase in the equilibrium state, which is critical for experimental exploration of exotic quantum effects associated with pseudospin-1 fermions, such as super Klein tunneling with almost 100% transmission probability for large incident angles,[37,38] and super collimation, which means guided unidirectional transport in the presence of a periodic potential, regardless of its initial direction of motion. Monolayer $K_2O$ may provide a promising platform for observing these fascinating phenomena.

It is well known that the physical properties of 2D materials can be effectively tuned by strain, like the bandgap engineering of 2H-TMDC semiconductors.[39] Strain has also been proposed to effectively modulate dielectric properties,[40] spin–orbit coupling,[41] thermal conductivity[42] and interlayer coupling (and mismatch) in vdW heterostructures[43] in 2D crystals. In the following, we will show that strain can induce interesting multiple topological phase transitions in the ML DMMOs. Before that, we introduce the in-plane stiffness constant, defined as $C = \frac{1}{S_0}\frac{\partial^2 E_s}{\partial \varepsilon^2}$, where $S_0$ represents the area of the unstrained cell and $E_S$ is the energy difference between unstrained and strained systems. Our calculation shows that the stiffness constants for $Na_2O$ and $K_2O$ are about 57.4 and 38.9 N m$^{-1}$, respectively. These values are significantly smaller than archetypal 2D systems of graphene ($\approx 340$ N m$^{-1}$)[44] and $MoS_2$ ($\approx 180$ N m$^{-1}$),[45] indicating that the properties of ML DMMOs can be readily manipulated by external strain. Strain stress curves are depicted in Figure S4 in the Supporting Information, which gives a maximum stress of about 2 GPa within ±10% biaxial strain. If taking the effective thickness as 2.85 Å, the instant maximum stress is 0.57 N m$^{-1}$.

We first show the changes in electronic band structure by applying biaxial strain. For $Na_2O$, multiple quantum phases can be generated, as shown in **Figure 3**. With a large compressive strain (for example −7% in Figure 3a), the double Weyl semimetal of $Na_2O$ becomes metallic with six extra Dirac points (due to historical reasons, spin–orbit-free Weyl points in 2D are also referred to as Dirac points[46]) emerging along the $\Gamma$-M axes apart from the original point at $\Gamma$. These new Dirac points are protected by three vertical mirror planes, while the band crossing along $\Gamma$-K is gapped without symmetry protection. When applying tensile strain, the energy difference between $E'$ and $A_1'$ decreases, leading to accidental band touching at $\Gamma$ between three bands at $\approx 3.8\%$ strain. After this topological phase transition, $Na_2O$ becomes a pseudospin-1 metal like pristine ML $K_2O$. Further increase in tensile strain causes band inversion between $E'$ and $A_1'$, and drives the system into a semiconductor, as shown in Figure 3c. We have plotted the complete quantum phase versus strain diagram of $Na_2O$ in Figure 3d. The results for ML $K_2O$ under biaxial strain are similar to $Na_2O$, which is summarized in the Supporting Information.

Next, we consider the effects of uniaxial strain, which has been experimentally employed in modifying the physical properties of graphene and TMDCs.[47–49] Unlike biaxial strain, uniaxial strain changes the crystalline symmetry of DMMOs by explicitly breaking the threefold rotation symmetry and vertical mirror symmetry. For uniaxial strain applied along the mirror plane, the $D_{3h}$ symmetry of DMMOs is reduced to $C_{2v}$. The uniaxial strain effect on band structures is pronounced for the double Weyl semimetal phase in $Na_2O$, as shown in **Figure 4**. It is distinctive that due to the symmetry reduction, the original double Weyl point (corresponding to the $E'$ doublet) is destroyed. Instead, a new pair of linearly dispersed Dirac points, as required by time reversal symmetry, emerge along $\Gamma$-M (see Figure 4a). Such a phase transition is consistent with the physical picture that a double Weyl point can be regarded as a superimposition of two single Weyl (Dirac) points, such as the case of bilayer graphene. To describe this strain-induced topological phase transition, we incorporate the effect of such uniaxial strain in the original Hamiltonian $H_{DW}$ by a perturbation term $H_s$ constrained by the remaining symmetry, namely

$$H(\boldsymbol{k}) = H_{DW}(\boldsymbol{k}) + H_s \quad (4)$$

with

$$H_s = -D\sigma_x \quad (5)$$

where we only keep the leading order term in $H_s$. Indeed, the double Weyl (massive Dirac) point at $\Gamma$ is splitting into two single Dirac points located at $\boldsymbol{k}_D^{\pm} = (\pm\sqrt{D/B_-}, 0)$ on the $k_x$ axis (i.e., $\Gamma$-M). Expanding the model at $\boldsymbol{k}_D^{\pm}$ leads to 2D Dirac model

$$H_D(\boldsymbol{q}) = \pm(v_+q_x\sigma_0 + v_-q_x\sigma_x + v_-q_y\sigma_y) \quad (6)$$

where the wave vector $\boldsymbol{q}$ is measured from $\boldsymbol{k}_D^{\pm}$, ± corresponds to the two Dirac points, and $v_{\pm} = 2B_{\pm}\sqrt{D/B_-}$. As we already mentioned above, the double Weyl point is characterized by a $2\pi$ Berry phase, whereas a single Dirac point corresponds to a $\pi$ geometric phase. At low-temperature, a $2\pi$ Berry phase is manifested by weak localization in quantum transport, whereas a π Berry phase leads to weak antilocalization. Thus, the uniaxial strain can induce an interesting transition from weak localization to weak antilocalization in ML DMMOs, which would be detectable by charge transport measurements. Due to the reduced symmetry, the aforementioned Dirac model also has a finite energy tilt term of $\pm v_+ q_x \sigma_0$, which may induce interesting squeezing effects on the Laudau level spectrum in magnetic field.[50] It is noteworthy that, in this phase, we will get a 2D topological insulator state when SOC is included, as also shown by the orange curve in Figure 4a. Although the energy gap is not significant (≈8 meV), we find that the corresponding topological $Z_2$ index is one, indicating a nontrivial 2D topological insulator (TI) state. The resulting topological edge state is plotted in Figure S5 in the Supporting Information, which also shows the evolution of Wannier charge center by means of WannierTools.[51] By constructing certain vdW heterostructures, we should be able to further enhance the SOC gap in Na$_2$O ML, making ML DMMOs promising 2D TI candidates for studying quantum spin Hall states.

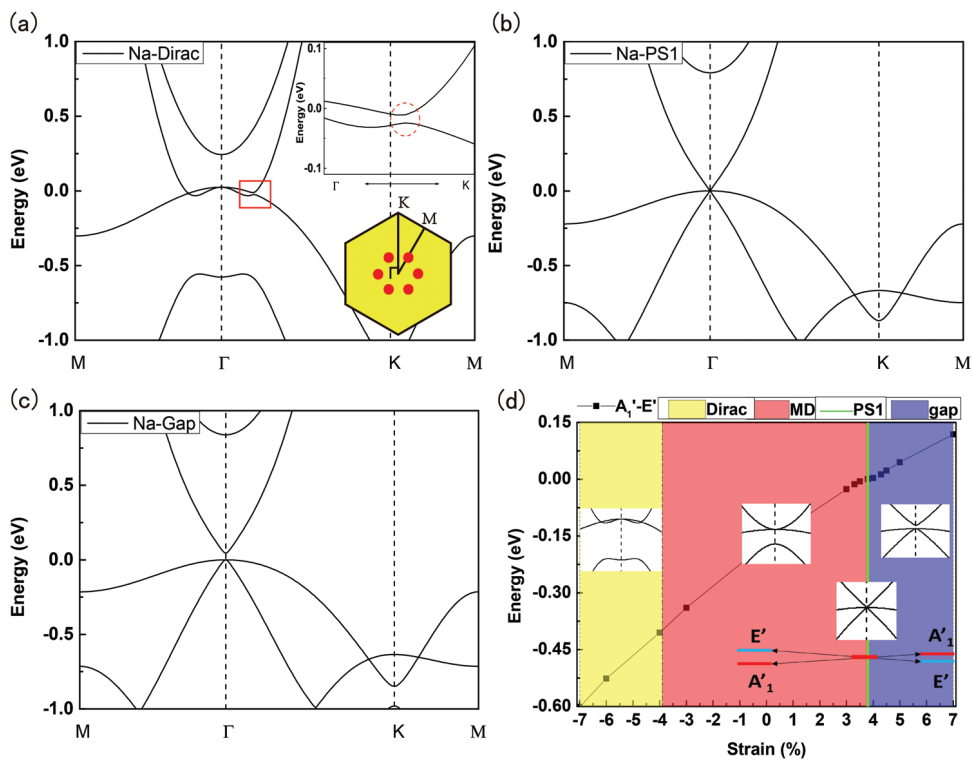

**Figure 3.** Band structures of ML Na$_2$O under biaxial strain of a) −7%, b) 3.8%, and c) 5%. Inset of (a) shows the emergence of six symmetry protected single Dirac points (red dots) along Γ-M paths. d) The complete strain versus topological phase diagram of ML Na$_2$O. The solid line represents the energy difference between the $A_1'$ and $E'$ representations. Four topological quantum states can be reversibly tuned by biaxial strain in ML Na$_2$O.

It is well known that TMDCs exhibits 1T or 2H phase under different environment. Similarly, ML DMMOs also have 1T phase counterparts which are dynamically stable.[52] For example, 1T-Na$_2$O could grow during the oxidation of Na (110)

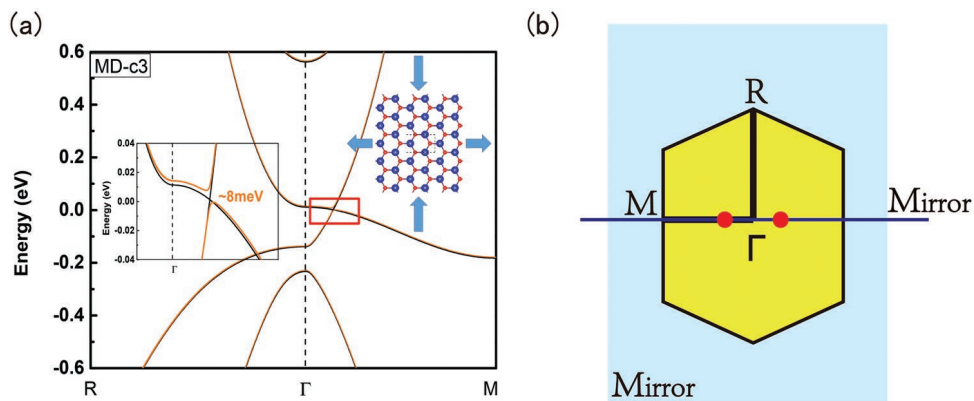

**Figure 4.** a) Band structure of ML Na$_2$O under uniaxial-strain of −3% applied along the mirror plane. When SOC is included, a nontrivial gap of ≈8 meV is determined (the orange curve). b) Schematic of the location of symmetry protected massless Dirac points (red dots) induced by uniaxial strain.

**1901939** (5 of 7)

surface[53] and 1T-Cs$_2$O may also be exfoliated from bulk.[54] Comparing to 1T-DMMOs, 2H-DMMOs are metastable phase, e.g., 2H-K$_2$O is ≈0.45 eV higher per unit in energy than 1T-K$_2$O (≈0.65 eV for Na$_2$O). However, these energy differences are much smaller than the energy difference between 1T and 2H phases of ML MoS$_2$ (≈0.85 eV). Since the metastable 1T phase of ML MoS$_2$ (not dynamically stable in theory[55]) has been already obtained in experiment[56] and the phase transition between 1T-MoS$_2$ and 2H-MoS$_2$ can be controlled by charge doping,[57] it is expected that 2H-DMMOs can also be obtained in experiments (see Figure S6, Supporting Information). We also suggest that the proposed two exemplary 2D materials of ML Na$_2$O and ML K$_2$O are likely to be synthesized by molecular beam epitaxy (MBE) or by chemical vapor deposition method. Another possible method is using alkali thin substrate with adsorption of oxygen, similar to the surface selenization to get reversible transition between 2H and 1T PtSe$_2$.[58] Although alkali compounds are not stable by reacting with ambient moisture when exposed to the air environment, we can encapsulate DMMO monolayers by protection layers of chalcogen elements, nonreactive oxides, or 2D insulating materials (such as boron nitride (BN)) in the glove-box environment before the device fabrication process. In situ synthesis and characterization techniques are also readily available and dialkali-metal monoxides can be grown in ultrahigh vacuum environment by MBE method, probed subsequently by scanning tunnel microscope and angle-resolved photoemission spectroscopy (ARPES) to determine the physical properties of these materials. For tuning topological states, strain can be introduced in ML DMMOs by transferring the thin films to flexible substrates or by using piezoelectric substrates. The induced topological phase transitions can be probed by scanning probe spectroscopy, ARPES, and ultimately by device fabrications and charge transport measurements.

## 3. Conclusions

In conclusion, we propose a family of new 2D materials—the monolayer dialkali-metal monoxides with the 2H-TMDC structure. These materials enjoy good stability and excellent flexibility. Most importantly, we find that these materials are novel types of 2D topological metals. In the equilibrium state, Na$_2$O is a 2D double Weyl semimetal, and K$_2$O is a 2D pseudospin-1 metal. They may host a range of fascinating physical effects such as universal optical absorbance, super Klein tunneling, and super collimation effect. Furthermore, rich topological phase transitions can be achieved in these materials by strain. Under biaxial strain, a series of transitions can be realized. Under uniaxial strain, the double Weyl point can be split into a pair of single Weyl (Dirac) points, accompanied with the transition from weak localization to weak antilocalization in quantum transport property. In addition, this phase is a 2D topological insulator state when SOC is included. Our work provides a new 2D material platform to explore new topological emergent fermions and their interesting effects, which is of significant fundamental importance. Besides, the exotic physical properties and effects associated with these fermions are expected to make the materials promising candidates for electronic device applications. The strain-induced metal–insulator quantum phase transition also makes these materials promising for making sensitive mechanical sensors.

## 4. Experimental Section

The first-principles computations were based on the DFT, performed by implementing the projector-augmented wave[59] method in the Vienna ab initio simulation package.[60] The generalized gradient approximation by Perdew, Burke, and Ernzerhof[61] was used to extract the exchange-correlation functional. A 20 Å vacuum layer was employed for all structures to avoid unphysical interaction between periodic images. Force and energy convergence criterion was set to 0.01 eV Å$^{-1}$ and 10$^{-6}$ eV, respectively. Energy cutoff was set to be 500 eV and the Brillouin zone was sampled using a 23 × 23 × 1 Γ-centered k-point mesh. Phonon spectra were calculated based on a 4 × 4 × 1 supercell by the finite displacement method using the PHONOPY package.[62] For Na$_2$O and K$_2$O, which are the focus of the work, the effect of SOC is negligible, and hence the results without SOC are used for the discussion unless specified (see the Supporting Information).

## Supporting Information

Supporting Information is available from the Wiley Online Library or from the author.

## Acknowledgements

C.H. and S.L. contributed equally to this work. This work was supported by the National Natural Science Foundation of China (Grant No. 61574123 and 11774305), Singapore Ministry of Education AcRF Tier 2 (MOE2015-T2-2-144 and MOE2017-T2-2-108), the National Key R&D Program of the Ministry of Science and Technology (MOST) of China (Grant Nos. 2016YFA0300204, 2017YFA0303002, and 2016YFA0300402), Zhejiang Provincial Natural Science Foundation (D19A040001), the Fundamental Research Funds for the Central Universities, and the 2DMOST, Shenzhen Univ. (Grant No. 2018028).

## Conflict of Interest

The authors declare no conflict of interest.